\documentclass[referee]{aa} 
\usepackage{graphicx}

\begin{document}
\title{15 colour photometry of Landolt SA95 standard star field
\thanks{The work is partly supported by Chinese National Sciences 
        Foundation under the contract No.19833020 and No.19503003}}
\author{Xu Zhou\inst{1},
	Zhaoji Jiang\inst{1},
	Jun Ma\inst{1},
	Suijian Xue\inst{1},
	Hong Wu\inst{1},
	Jiansheng Chen\inst{1},
	Jin Zhu\inst{1},
	Weihsin Sun\inst{2},
	Rogier A. Windhorst\inst{3}}

\offprints{Xu ZHOU}

\institute{National Astronomical Observatories, Chinese
              Academy of Sciences, Beijing 100012, P. R. China\\
              \email{zhouxu@bac.pku.edu.cn}.
             \and
              Institute of Astronomy, National Central University, 
              Chung-Li 32054, Taiwan.
             \and
              Department of Physics and Astronomy, Box 871504, 
              Arizona State University, Tempe, AZ 85287-1504}

\date{Received; accepted}

\abstract{In this paper, we present a  set of photometric
   observations in 15 colors of stars in the Landolt
   SA95 field with the BATC system. The wavelengths 
covered by the system range from 300 nm to 1000 nm. 
Visual magnitudes of the stars being studied in the field are 
from 10th to 20th mag.  The observational methodology and 
the data reduction procedures are described.  The relationships
between the BATC intermediate-band system and the Landolt $UBVRI$ 
broad band system are obtained. A catalogue of the photometry has been 
produced which contains the SEDs of 3613 stars.  The electronic form 
of this catalogue can be accessed at the CDS 
via anonymous ftp to cdsarc.u-strasbg.fr.

   }

\authorrunning{ZHOU Xu et al.}
 \maketitle
%
\section{Introduction}

   Landolt (1983) gave his measurements of $UBVRI$ magnitudes of 
      233 standard stars in a strip about two degrees wide centered
     on the celestial
   equator for the purpose of homogeneous flux calibration from 
both hemispheres. 
     Further more, Landolt (1992) presented the $UBVRI$ photometry
of another 298 standard stars around the celestial equator,
     within a visual magnitude range of 11.5 -- 16.0 and a $B-V$ colour
     range from $-0.3$ to $+2.3$.
     The Landolt catalogues have been widely
     used by observers using intermediate to large size
   telescopes. By using the Landolt standards catalogues,
   Galad\'{\i}-Enr\'{\i}quez et al. (2000) obtained $UBVRI$ photometry of
     11 different fields around Landolt standards and gave 
681 secondary standards with a visual magnitude range from 9.7
   to 19.4 and a $B-V$ range from 1.15 to 1.97.

     With the capability of imaging large fields, 
   Beijing-Arizona-Taipei-Connecticut (BATC) survey program 
has observed a number of selected fields which include 
  Landolt SA95 field centred at
     $\alpha = \rm{3^{h}54^{m}17^{s}}$,
   $\delta= +00^\circ19^{\prime}08^{\prime\prime}{\mbox{}\hspace{-0.1cm}}$
   (2000.0).  So far Landolt SA95 has been one of the 
fields with higher observing quality.  Thus we select it to demonstrate the 
photometric quality of the BATC system.  Standard BATC data reduction
processes have been applied to this field.  
In this field, 7 stars have been measured in Landolt (1983),
   and 45 are listed in the catalogue of Landolt (1992). In addition,
   7 stars in this field were included in the secondary standard catalogue of
   Galad\'{\i}-Enr\'{\i}quez et al. (2000).

   The main purpose of this paper is to present the spectral energy distributions
   (SEDs) of the field stars in the format of 15 colours, and to show 
 the relationships between the BATC  and the
   $UBVRI$ photometric systems derived 
   by using the stars in  the catalogues
   of Landolt (1983), Landolt (1992), and Galad\'{\i}-Enr\'{\i}quez et
   al. (2000). In Sect. 2, we present the observing processes of the Landolt SA95
   field.  In Sect. 3, we describe  the method of
   data reduction. The magnitude error of observation and data
   reduction are discussed in Sect. 4. The discussion of the system transformation
is given in Sect. 5.  Sect. 6 gives the conclusions of this study.

%
\section{Observation}

{\bf Large field multi-colour photometry has been and is still being}
conducted 
with the BATC photometric system.   The major observing facility employed
is the 60/90 cm f/3 Schmidt Telescope located at Xinglong Station of 
 Beijing Astronomical Observatory (BAO).   A Ford Aerospace $2048 \times
   2048$ CCD camera with 15 micron pixel size is mounted at the
   Schmidt focus of the telescope. The field of view of the CCD is
   $58'\times58'$ with a plate scale of 1.7 arc-second per pixel.  An image of 
the Landolt SA95 field taken with this system in $i$ band (666 nm) is shown
      in Fig. 1.


   The BATC filter system includes 15 intermediate band
   filters, covering a range in optical wavelengths from 300 to
   1000nm (Fan et al. 1996; Yan et al. 1999; Zhou et al. 2001).
   The filters are designed specifically to avoid contamination from
   the brightest and most variable night sky emission lines. The
transmission curves of the 15 filters are given in Fig. 2.


   The BATC photometric system defines the magnitude zero points 
in a way similar to the
   spectrophotometric AB magnitude system. The AB system is a
   monochromatic $\widetilde{f_{\nu}}$
     system first introduced by Oke \& Gunn (1983), based upon
   the spectral energy 
   distributions (SEDs) of the four F sub-dwarfs,  HD19445, 
   HD84937, BD+262606, and BD+17 4708.

   A great advantage of the AB magnitude system is that the
   magnitude is directly related to physical units.
      As in the definition of AB
   magnitude system, the BATC magnitude system is defined as follows:

   \begin{equation}
    m_{\rm batc} = -2.5\cdot {\rm log}\widetilde{F_{\nu}} - 48.60,
   \end{equation}
   where $\widetilde{F_{\nu}}$ is the flux per unit frequency in unit
   of erg s$^{-1}$ cm$^{-2}$ Hz$^{-1}$.

   In the BATC system (Yan et al. 1999), $\widetilde{F_{\nu}}$ is defined
   as 
   
   \begin{equation} 
     \widetilde{F_{\nu}}=\frac{\int{d} ({\rm log}\nu)f_{\nu}R_{\nu}}
      {\int{d} ({\rm log}\nu)R_{\nu}},
   \end{equation}
   which ties the magnitude to the number of photons detected
   by the CCD rather than to the input flux (Fukugita et
   al. 1996). The response of the system $R_{\lambda}$, which is 
   used to relate $f_{\nu}$ and $\widetilde{F_{\nu}}$,  includes only
   the filter transmissions. Other effects such as the quantum
   efficiency of the CCD, the response of the telescope's optics, and the
   transmission of atmosphere, etc., are ignored. This makes the BATC
   system filter-defined, since the bandwidths are intermediate in
   size and all the responses are essentially flat within each 
   specified passband.

   The magnitudes of the 4 Oke \& Gunn (1983) standards have 
subsequently been refined
   by many authors (Oke 1990;  Castelli \& Kurucz 1994), with 
   Fukugita et al. (1996) presenting the latest
   re-calibrated fluxes of these four standards.
      Their magnitudes in the BATC system have been slightly corrected 
      recently using the data obtained in a number of 
photometric nights (Zhou et al. 2001).

 In order to understand the sensitivity curve of the BATC 
   photometric system, we run a test observation of HD84937 in 
   a night of good observing quality (Zhou et al, 1999). With equal exposure 
   time, we obtained
   the images in 13 BATC filter bands. The differences between instrumental 
   magnitude
   and BATC magnitudes of this star can be regarded as the sepectral
   sensitivity of the BATC photometric system. The result is shown in 
   Fig. 3. The magnitude is normalized to $i$ (666 nm) band. The spectral
   sensitivity curve includes all effects of the system such as the 
   CCD response, filter band width and its transmissions, and the 
   {\bf transparency} and reflection efficiency of telescope optics, and
   the spectral extinction of the atmosphere.
   From this figure, we can see immediately that the $i$ (666 nm)
   band is the most sensitive one in the BATC photometric system, while the
   $b$ (389 nm) and $p$ (975 nm) bands are about 4 magnitudes less sensitive
   than the $i$ band.

   \begin{figure}
   \centering
   \includegraphics[angle=-90,width=88mm]{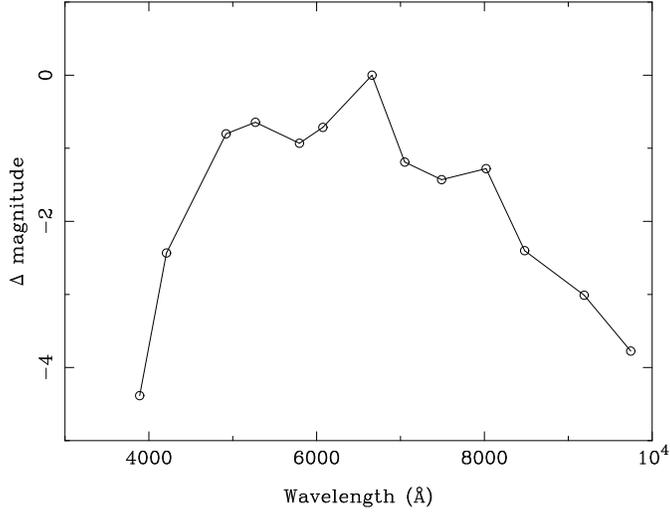}
      \caption{The spectral sensitivity curve of the BATC 
               photometric system obtained from the observations of 
               standard star HD84937 in 13 bands. The magnitude
               difference is normalized to $i$ band (666 nm)}
   \end{figure}

   We have observed the Landolt SA95 field in a total of 41 nights
in the period from December 13, 1994 to December 16, 1999. 
A total of 189 high quality images were selected for further 
photometric analysis. The total
   exposure time and image numbers of each filter band are listed in
   Table 1.  Typical long exposure time lasts for 10 to 20 minutes, while 
   it is 5 minutes for short exposure. The short exposure images were
   taken during photometric nights for flux calibration.


   \begin{table}
      \caption[]{Statistic table of observation.}
         \begin{tabular}{ccccccc}\hline
            \hline
           No. & Filter & Wavelength & Total Exp. &
           Number of & Number of & Calib. \\
                &         & (nm)      & (hour)    &
           images & calibrations & error\\
           \hline
             1 & $a$ & 337.15 & 2:17 &  6 &  2 &  0.001 \\
             2 & $b$ & 390.69 & 5.50 & 34 &  4 &  0.033 \\
             3 & $c$ & 419.35 & 0.96 &  7 &  2 &  0.000 \\
             4 & $d$ & 454.00 & 1.33 &  9 &  4 &  0.007 \\
             5 & $e$ & 492.50 & 2.30 &  8 &  2 &  0.010 \\
             6 & $f$ & 526.68 & 1.22 &  6 &  7 &  0.007 \\
             7 & $g$ & 578.99 & 2.23 & 10 &  8 &  0.007 \\
             8 & $h$ & 607.39 & 1.33 &  8 &  7 &  0.007 \\
             9 & $i$ & 665.59 & 1.75 & 11 & 11 &  0.006 \\
            10 & $j$ & 705.74 & 4.22 & 16 &  9 &  0.004 \\
            11 & $k$ & 754.63 & 3.30 & 13 &  6 &  0.008 \\
            12 & $m$ & 802.32 & 3.69 & 17 &  6 &  0.006 \\
            13 & $n$ & 848.43 & 3.73 & 12 &  7 &  0.004 \\
            14 & $o$ & 918.22 & 4.40 & 14 &  3 &  0.006 \\
            15 & $p$ & 973.85 & 4.15 & 15 &  3 &  0.006 \\
	   \hline
         \end{tabular}
   \end{table}

   {\bf In the nights judged photometric by the observers,}
   the standard
   stars of  Oke \& Gunn (1983) were observed between air-masses 1.0 and
   2.0 for each programmed filter band. The standards were observed in
   short exposure to avoid saturation.
   Normally, there is enough time to obtain flux calibration images for 
 4 to 6 filters in each photometric night.  In each image, the standard star is 
located at or near the center.  
  
      A subsection of $300 \times 300$ pixels on CCD was
   used in these standard star observations in order to save readout time
   and disk space. The extinction coefficients and magnitude zero
   points obtained from the standard star observations
 are then used to calibrate other BATC field images. The observations
   for calibration are described in detail in Zhou et al. (2001).

%
\section{ Data reduction}

 \subsection{Pipeline I}

   An automatic data reduction procedure called ``Pipeline I'' 
has been developed as a standard 
for the BATC multi-colour sky survey 
(Fan et al. 1996), which includes bias subtraction
   and flat-fielding of the CCD images.  The dome flat-fielded
   images in each filter were subsequently combined. 
 When combining the images, the cosmic ray hits and bad pixels
   were corrected through comparison of multiple images.  The reduction
and combination of Landolt SA95 images followed the same way.  
  
   Before the combination, the images were
   shifted and rotated according to the precise locations of 
stars in the HST Guide Star
   Catalogue (GSC) (Jenkner et al. 1990).  A plate solution of 8 
parameters were obtained by positioning the GSC
   stars of the images in each colour.  Then the positions of the stars in the
   images ($x,y$) can be easily transformed to the equatorial
   coordinates ($\alpha$ and $\delta$).  The final RMS errors in 
   positional accuracy of the stars are about 0.5 arc-second.

 \subsection{ Pipeline II}

   A ``Pipeline II'' program has been developed to measure the 
magnitudes of point sources in BATC images.  It is based on Stetson's
standard procedure of DAOPHOT (Stetson 1987). 
  The Pipeline II includes 4 major steps:

   1. to find all the sources in the images,

   2. to perform aperture photometry at 10 different radii of apertures,

   3. to construct PSF parameters and a look-up table with bright
    isolated stars, and 

   4. to perform PSF fitting to each point source and obtain its magnitude.

   In the first step, 
      objects are identified in the combined images. An object 
      detected at least in 3 bands is considered real.  Certain objects
that appear at the edge of some images are marked in the final output
SED catalogue.  
    
   In a series of test with different image profiles, 
we find that the analytical
   profile ``moffat25'' best fits the observed PSF in the BATC
   images. In a one degree field, the FWHM of PSF can change 0.2
   arc-seconds from one place to another.  This is due to the fact
that the Schmidt focal plane is curved while the CCD surface is flat 
 (Fan et al. 1996).

      To obtain quasi-linear variation for the PSF lookup table, 
      we divided each image into $3\times3$ sections with some 
      overlap between adjacent sections.  In each such section, we
normally select more than
   20 bright, isolated stars for PSF fitting with ``moffat25''
   profile, and obtain a look-up table to trace the linear residual
   of the PSF's across the sub-images.

    In our automatic procedures of data reduction, most of the
parameters in each step were optimised and fixed, leaving only
a few, such as the FWHM of the object, to vary according to 
the seeing condition at the time the image was taken.  
   
It is difficult to obtain good PSF fitting for faint stars due to
their asymmetry in profile.  However, in order to construct
an SED catalogue as complete as possible, faint stars were 
measured with aperture photometry using an aperture with a 
diameter of 9 pixels.   The difference between PSF fitting
   and aperture photometry is corrected using the
bright, isolate stars in the same field.  
These faint stars are marked in the catalogue.

   At the completion of photometry, 
the SEDs of all the measurable objects are obtained.  
All of these objects will have also been visually inspected.  There 
are usually some special objects, 
such as small extended sources, unresolved blended stars, and 
ghost images around very bright stars, etc. 

{\bf These special objects are marked, along with the objects close to
saturated stars.}

 \subsection{Calibration}

   Using the images of the
   standard stars observed in the photometric nights,
   we derive iteratively the extinction curves, and the slight
   variation of the extinction coefficients with time (Zhou et al.,
   2001). The extinction coefficients at any given time in a night
   $[K + \Delta K (UT)]$  and the zero point of the 
instrumental magnitude ($C$) were  obtained.

   The instrumental magnitudes ($M_{\rm inst}$) of selected bright,
 isolate stars can be readily transformed to the 
BATC AB magnitude system ($M_{\rm batc}$) by

   \begin{equation}
       M_{\rm batc} = M_{\rm inst} + [K + \Delta K(UT)]\cdot \chi + C,
   \end{equation}
   where $\chi$ is the airmass at which the image was taken, and $C$ is the
   zero point of instrumental magnitude.

   These bright stars, of which the calibrated magnitudes were obtained
   in short exposures, were taken as secondary standards and were
subsequently used to tranform the magnitudes in the process of 
calibration.  Normally about 30 stars, 
which are brightest, isolated, and unsaturated,
 were selected to perform calibration in the combined image of each filter. 
 
       The error in magnitude for a single star is about 0.03 mag, 
       a value obtained from the comparison of the magnitudes in 
       a short exposure with that from the combined image. The 
  mean error propagated using 30 stars should thus be less 
       than 0.01 mag.

   The flux calibration for the images taken in 
   the 18 {\bf photometric nights has been done}  several times. 

 We then take the mean value of calibration
   and obtain a final calibration error for each filter. Table 1 lists
   the times that calibration has been performed 
and the calibration error for each
   filter. The error of calibration, $Error$, listed in the last column
   of Table 1 was calculated using 

   \begin{equation}
      Error = \sqrt{\frac{ \Sigma (mag-\overline{mag})^2}{n (n-1)}}.
   \end{equation}

   In most filters, the calibration errors are equal to or less
   than 0.01 mag. The errors in $a$ and $b$ bands (337 and 390 nm) are about
   0.02 mag, larger than that in the other bands. This may come from the 
fact that the seeing effect is worse in shorter wavelengths, 
and the bright stars are not
so bright as in other bands due to the lower intensity of the stars 
as well as the lower sensitivity of the CCD in 
shorter wavelengths.  

{\bf A naive guess will predict that the error in $a$ band 
should be larger than that in $b$, but the opposite was observed. 
This may be due to small number statistics.} The
        same effect may also present in the errors of certain bands with a
        small number of calibrations.  
   In general, except for $a$ and $b$ bands, the total 
        calibration error is about 0.01 mag.

   The final SED catalogue of the Landolt SA95 field is created also in 
electronic format.  In this catalogue, each star takes up 2 lines. 
The first line contains the 
   coordinates and magnitudes of the stars in 15 bands. The second line
   gives 
   {\bf the estimated error of the magnitudes.}
 For the stars also found in
   the USNO all-sky catalogue, their USNO names and B and R magnitudes 
   are presented in
   the beginning of the second line. There are a total of 3613 stars
in the catalogue, listed in order of increasing right ascension.
      
The catalogue of photometry is available in electronic form at 
      the Centre de Donn\'ees astronomiques de Strasbourg (CDS) 
      via anonymous ftp to cdsarc.u-strasbg.fr (130.79.128.5).

\section{The total measurement error}
 \subsection{The sources of magnitude error}

   The total error of calibration comes from the 
following sources: a) bias
   (over-scan) correction and CCD readout noise (0.001 mag); b)
   flat-field correction (0.003 mag); c) photometry of PSF fitting
   (0.02 mag); d) calibration error (0.01 mag).

      We employed the Lick Observatory data-taking system, which 
      automatically subtracts the overscan of each image during readout, 
      and records the overscan in the last column. We 
      then process each program image through median filtering 
      the original overscan and adding 
      back to the image the difference between the original 
      subtracted overscan and this filtered overscan. In this 
      way, the signal-to-noise ratio (S/N) of the overscan 
      is increased and any residual large-scale pattern produced 
      in the image by imperfections in the overscan is removed.
      The overscan will increase the noise value of the image 
      background. The magnitude error arising from the bias 
      depends on the brightness of a star, and it can be treated
      as the background error in photometry.

   Normally, 12 dome flat-field images in each band were taken within
   24 hours before and/or after the observations. To avoid 
   {\bf shutter effects,}
   the exposure time is always longer than 2 minutes. 

The number of ADU of
   each flat-field image is higher than 20000. We estimate the
   error of the flat-field correction to be 0.3\%
   (Fan et al. 1996; Zhou et
   al. 2002).  When the statistic error is considered, the 
   flat-field error in the Landolt SA95 field is about 0.003 mag.

   By comparison with different flux calibrations, the 
calibration error is about 0.01 mag. 
    Thus the main error in the magnitude in the catalogue is the photometric
       error of PSF fitting. It comes from the statistical photon counting error 
      and the sky background noise. The value of these photometric
       measurement error is dependent on the brightness of the objects. 
   In our catalogue,  a magnitude-dependent
 error is given for each object. For brighter stars, 
the errors are about 0.02 mag.  As to the fainter stars, the magnitude
   errors are relatively larger depending on their magnitudes.

 \subsection{ Test of total measurement error}

   In order to test the error estimate 
of our photometry, we made an observation of
   the Landolt SA95 field in $i$ band (666 nm). 
As shown in Fig. 4, the test 
   images were taken at positions along the 
top edge aligned with the top two corners of the
  Landolt SA95 field.
   Through the same data reduction processing, we obtained the
   magnitudes of the objects in the test images. The photometric error
   is obtained by comparing the magnitudes of the objects in the
   overlapping area of the test and the combined 
Landolt SA95 images. Figs. 5 and 6 
   show this comparison. The circles in Fig. 5 
   represent the real magnitude
   difference between the test image and the Landolt SA95 image, and
   the plus signs in Fig. 6 represent the magnitude errors given in the
   catalogue of the test image. From these two figures, it is apparent 
that the errors in the catalogue show the same 
   distribution as the real measurement errors.

%
   \begin{figure}
   \includegraphics[angle=-90,width=120mm]{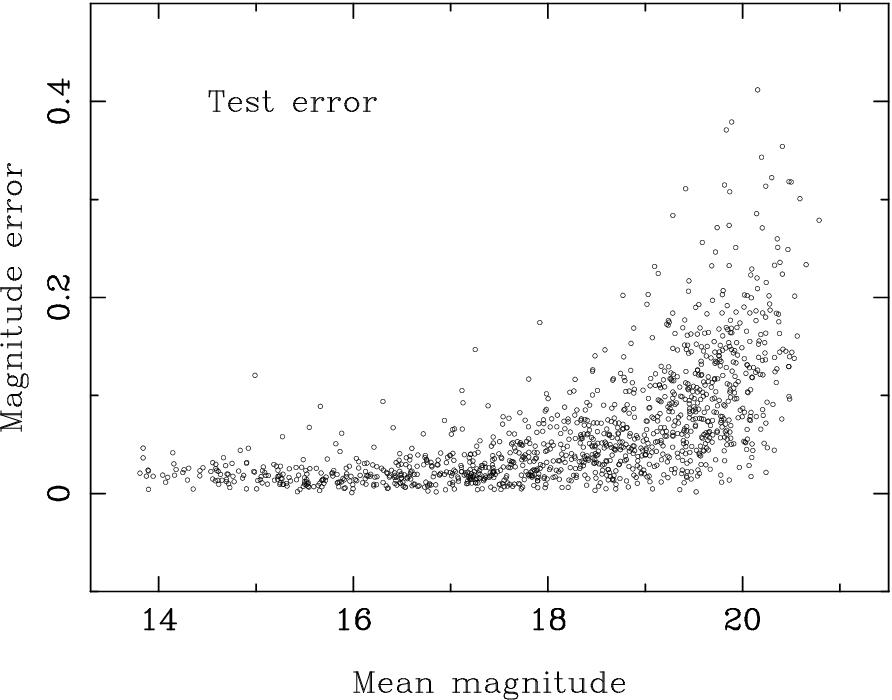}
  \caption{The
   diagram of magnitude error versus magnitude in $i$ band
   (666nm). The error is the magnitude differences between the magnitude 
    obtained from the test images and the combined image of SA95 field.}
   \includegraphics[angle=-90,width=120mm]{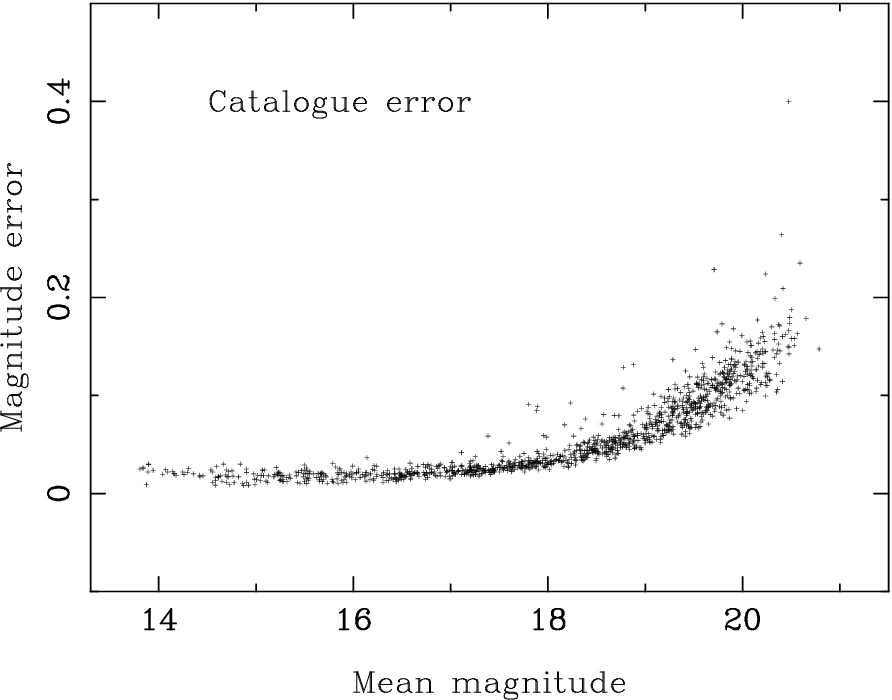}
   \caption{The
   diagram of magnitude error versus magnitude in $i$ band
   (666nm). The magnitude error is the measurement error listed  in the
   catalogue.}
   \end{figure}

\subsection{Magnitude limit and completeness of the catalogue}

    Two concepts have been adopted to define the quality
of the measurements in the resulting catalogue.  The first 
one is the limiting magnitude of the
      observation,  and the other one 
is the completeness of the catalogue. The
      limiting magnitude is {\bf defined} as the magnitude at which the mean
      magnitude error of the star becomes 0.1 mag.  By fitting the
      distribution of the error bars along the magnitudes, we obtained the
      curve of the error bars as a  function of the magnitude,
   
      \begin{equation}
         \Delta mag = a_{1} 10^{(mag/2.5)} + a_{0} ,
      \end{equation}
      where  $a_{0}$ and $a_{1}$ are constants obtained in the fitting process. 

      From the curve we obtain the limiting magnitude for each band and the 
error of which is 0.1 mag.  The completeness of the
      catalogue, which is also expressed as a magnitude, is defined
      as the maximum of the star counts as a function 
      of the magnitude. It shows how many stars can be detected 
brighter than certain magnitude limit, which gives an estimate of the 
completeness of the sample.  
      We have plotted the histogram of magnitudes for each filter
      band with binning width of 0.5 mag. The maximum of each 
      magnitude bin is taken to be the indicator of the completeness.

      The results of limiting magnitude and the estimated completeness are shown
      in Table 2.  In this table, we can see that the limiting
      magnitudes of the $a$ and $p$ filter bands are about 2 mag
      brighter than the $i$ band, even though the exposure time 
in these two bands is much
longer than that in the $i$ band. Another interesting point from
the table is that the difference in
      the completeness among various bands 
is less than that of the limiting 
      magnitude. The reason is
      that aperture photometry was performed to measure the magnitude of faint 
      objects whose PSF's were difficult to get.  By doing so, we have obtained
the magnitude estimates for the majority of the stars in the field and the number
of objects missed by both methods should be insignificant.  

   \begin{table}
      \caption[]{Magnitude limit ($\Delta mag=0.1$) and completeness defined by 
the maxima in the magnitude histogram.}
      \vspace{0.5cm}
      \begin{tabular}{ccc}
      \hline
      \hline
       filter & limit of mag. & completeness of mag. \\ \hline
        $a$ & 18.9 & 20.5 \\
        $b$ & 19.6 & 21.0 \\
        $c$ & 19.1 & 20.5 \\
        $d$ & 19.8 & 21.0 \\
        $e$ & 20.5 & 22.0 \\
        $f$ & 20.2 & 21.0 \\
        $g$ & 20.2 & 20.5 \\
        $h$ & 19.9 & 21.0 \\
        $i$ & 20.3 & 20.5 \\
        $j$ & 20.3 & 20.0 \\
        $k$ & 19.8 & 19.5 \\
        $m$ & 19.0 & 19.5 \\
        $n$ & 18.6 & 19.0 \\
        $o$ & 18.6 & 19.0 \\
        $p$ & 17.6 & 18.5 \\
      \hline
      \end{tabular}
   \end{table}

\section{System conversion}

   In our previous work (Fan et al. 1996), we made a comparison of
   magnitudes between the BATC system and the $UBVRI$ system through the stars
   in the field of open star cluster M67 (Gilliland 1991).

   In the papers of Landolt (1983, 1992) and of
   Galad\'{\i}-Enr\'{\i}quez
       et al. 
   (2000), there exists $UBVRI$ {\bf  photometry for 55 stars} 
   in the field of Landolt SA95. By cross-checking the
   published catalogues  of the standards and our SED catalogue, we 
   found 48 stars in common, since the other 7 stars are too bright and 
   saturated in our images in most bands. 
       In Tables 4 and 5, there are 42 stars taken from SA95 field of
       Landolt (1992) which includes 4 stars of Landolt (1992) and 
       all the 7 stars in 3'rd field of Galadi-Enriquez et al. (2000).

   The coordinates of one of the stars, 95-285, may be wrong. At
     $\alpha=\rm{03^{h}55^{m}46^{s}},
     \delta=+00^\circ23^{\prime}40^{\prime\prime}$
   (2000.0), we did not find any objects. From the finding
   chart of Landolt (1992) and the given magnitude, we suggest that the object  
   of 95-285 should be located at
   $\rm{03^{h}55^{m}44\hspace{0.1cm}{.}\hspace{-0.15cm}^{s}1},
   +00^\circ25^{\prime}09^{\prime\prime}{\mbox{}\hspace{-0.1cm}.6}$
   (2000.0).

   By comparison of the $UBVRI$ and the BATC magnitudes of these 48 stars,
   we obtained the relationships between the two photometric systems
   via linear fitting method, which are presented in figures 7-11, respectively. 

   \begin{equation}
      m_{U} = m_{b} + 0.6801 (m_{a}-m_{b}) - 0.8982 \pm 0.143,
   \end{equation}
   \begin{equation}
      m_{B} = m_{d} + 0.2201 (m_{c}-m_{e}) + 0.1278 \pm 0.076,
   \end{equation}
   \begin{equation}
      m_{V} = m_{g} + 0.3292 (m_{f}-m_{h}) + 0.0476 \pm 0.027,
   \end{equation}
   \begin{equation}
      m_{R} = m_{i} + 0.1036 \pm 0.055,
   \end{equation}
   \begin{equation}
      m_{I} = m_{o} + 0.7190 (m_{n}-m_{p}) - 0.2994 \pm 0.064.
   \end{equation}

   \begin{figure}
   \includegraphics[angle=-90,width=100mm]{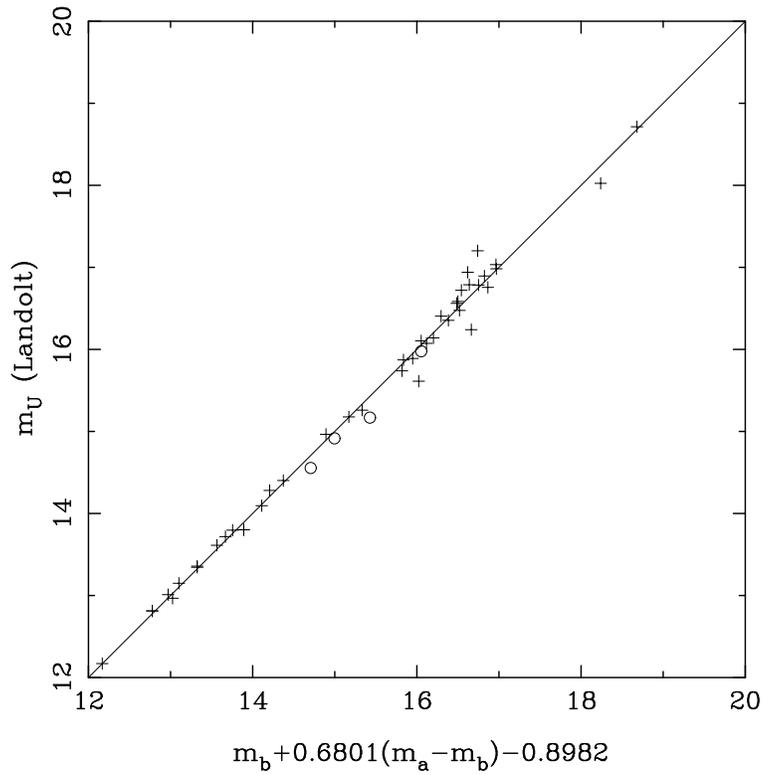}
   \caption{The comparison between Landolt $U$ magnitude and the BATC 
            $a$ and $b$ magnitudes. Here $b$ (391nm) band is the filter
            band nearest to $U$ band, and $a$ (337nm) and $b$ (391nm) 
            bands are used for colour correction.}
   \end{figure}

\begin{figure}
   \includegraphics[angle=-90,width=100mm]{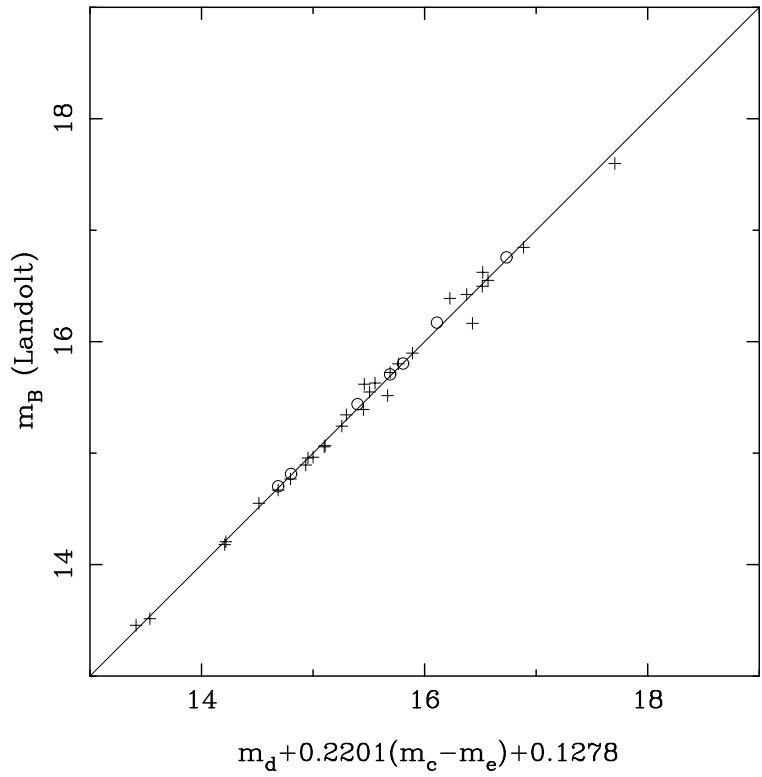}
   \caption{The comparison between Landolt $B$ magnitude and the 
            BATC $d$, $c$ and $e$ magnitudes. Here $d$ band (454nm) 
            is the filter band nearest to $B$ band, and $c$ (419nm) 
            and $h$ (493nm) bands are used for colour correction.}
   \end{figure}

   \begin{figure}
   \includegraphics[angle=-90,width=100mm]{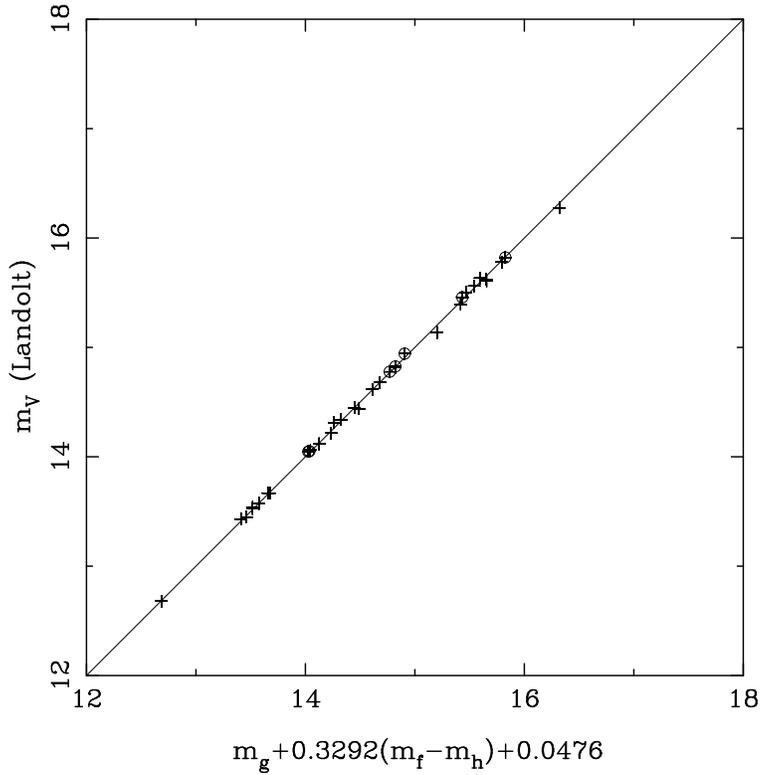}
   \caption{The comparison between Landolt $V$ magnitude and the BATC 
            $g$, $f$ and $h$ magnitudes. Here $g$ bands (580nm) is 
            the filter band nearest to $V$ band, and $g$ (580nm) and $h$ 
            (607nm) bands are used for colour correction.}
   \end{figure}

   \begin{figure}
   \includegraphics[angle=-90,width=100mm]{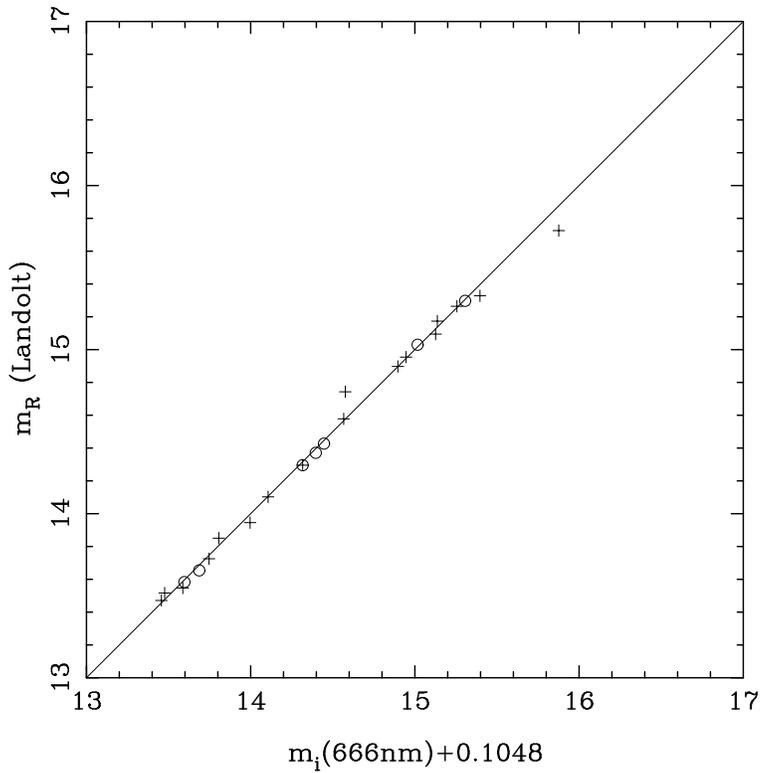}
   \caption{The comparison between Landolt $R$ magnitude and the BATC 
            $i$ (666nm) band. Colour correction does not improve the
            quality of the fit in this case.}
   \end{figure}

   \begin{figure}
   \includegraphics[angle=-90,width=100mm]{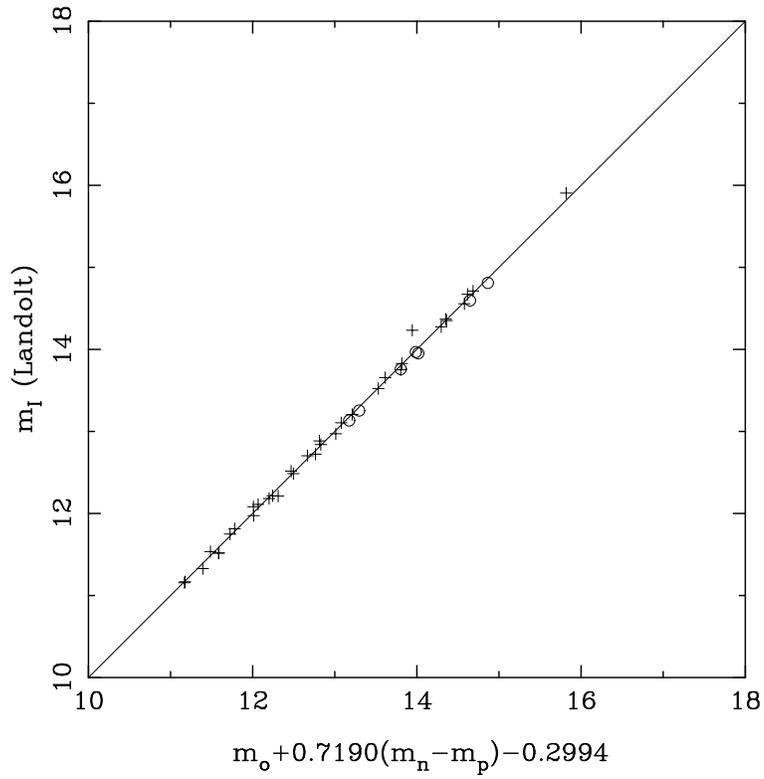}
   \caption{The comparison between Landolt $I$ magnitude and the BATC 
            $o$, $n$ and $p$ magnitudes. Here $o$ (918nm) band is nearest 
            to $I$ band, and $n$ (848nm) and  $p$ (974nm) bands are used 
            for colour correction.}
   \end{figure}

   In Figs. 7-11, we use different symbols for the points from 
   catalogues of Landolt (1983, 1992) and from  the catalogue of 
   Galad\'{\i}-Enr\'{\i}quez et al. (2000).  By careful inspection of 
the figures, we did not find any
systematic difference. 

   Some of the standards are too bright and are out of our 
magnitude ranges. We can thus only use part of
   the Landolt standards to perform fitting. Table 3 list useful 
   informations in the fitting process, which include  the number of
   stars used, the range of colour index $B-V$ and the errors 
   of the fitting in each band of the $UBVRI$ system.

   \begin{table}
      \caption[]{The fitting statistics of the $UBVRI$ from BATC
                 photometry. Col. 1 lists the name of the band. 
                 Col. 2 lists the number of the stars used for fitting.
                 Col. 3 and 4 list the ranges of color index of each band,
                 and the last column lists the RMS fitting errors.}
      \vspace{0.5cm}
      \begin{tabular}{ccccc}
      \hline
      \hline
       filter & No. of stars & $(B-V)_{min}$ & $(B-V)_{max}$ & Error\\
      \hline
        $U$ & 45 & $-0.22$ & 2.00 & 0.143 \\
        $B$ & 37 & $-0.22$ & 1.76 & 0.076 \\
        $V$ & 33 & $-0.22$ & 1.53 & 0.027 \\
        $R$ & 26 & $-0.22$ & 1.53 & 0.055 \\
        $I$ & 40 & $-0.22$ & 1.76 & 0.064 \\
      \hline
      \end{tabular}
   \end{table}

   The comparison and 
fitting between $V$ band and the BATC $g$, $f$ and $h$ bands
   presents the best case in this study, and 
   the RMS is about 0.027 mag. The RMSs of fitting for the other
   broadbands are somewhat larger.

   To obtain transformation equations for the derivation of
   approximate $UBVRI$ magnitudes and colours from BATC photometry is
   important as well as essential. The above relationships can be
   used for the system transformation and can also be applied to the
   verification of our internal estimation of accuracy by using
   Landolt's catalogues. Although the many comparisons and fits
   between the two systems are good and self-consistent, the study
   will not be limited to this single field. BATC survey program has
   been and will be observing more target fields including other
   standard fields. Additional comparisons in the future with more
   stars in larger colour range will provide better information as
   to how the magnitude measurements can be transformed among various
   observing systems.
    
\section{Conclusion}
   In a total of 41 nights in the period 
from December 13, 1994 to December 16, 1999, we made
   observations of the Landolt SA95 field. A total of 189 images with acceptable quality
 were selected for photometric measurements. A complete SED catalogue 
   in 15 colours in the BATC system for the stars in the field of
   Landolt SA95 is presented.  The wavelengths coverage with the 15
   intermediate filter bands are from 300 nm to 1000
   nm. Visual magnitude ranges from 10th to 20th mag. We describe the
   methods of observation and data reduction, and analysed the possible error 
   of our measurement. By comparison with Landolts $UBVRI$
   broad band photometric magnitudes of 48 stars, the relationships
   between the BATC intermediate-band system and Landolt $UBVRI$
   broad band photometic system are obtained.  A catalogue is created 
in which a total of 3613 stars is included.  The catalogue is also published in 
electronic form and is available at CDS ftp site to cdsarc.u-strasbg.fr (130.79.128.5).

\begin{acknowledgements}
We are indebted to the anonymous referee for many critical comments 
and helpful suggestions, and for English editing that have greatly
improved our paper.  We thank Yao-Hua Li
  for his management and support of the instruments.  We also thank
  the assistants who helped with the observations for their hard work
  and kind cooperation.  Wei-Hsin Sun acknowledges the support from National Science Council under 
the grant NSC 89-2112-M-008-021.  
\end{acknowledgements}

\onecolumn
{\tiny
   \begin{table}
   \caption{The positions and names of the SA95 standard stars used in
            system conversion. Col. 1 is the sequence number in increasing
            right arsension ($\alpha$). Col. 2 and 3 are the equatorial
            coordinates in epoch 2000.0. Last 3 columns give the names in 
            catalogues of Landolt (1983), Landolt (1992) and
            Galad\'{\i}-Enr\'{\i}quez et al. (2000).}
         \begin{tabular}{cccccc}\hline
           \hline
 Number &  $\alpha$(2000.0) & $\delta$(2000.0) & L-1992 & G-2000  & L-1983 \\
  (1) & (2)  & (3) & (4) & (5) & (6) \\
01 & 03:52:34.01 & --00:02:27.3 & & 3-1 & \\
02 & 03:52:36.96 & --00:03:32.5 & & 3-2 & \\
03 & 03:52:39.95 & --00:03:07.8 & & 3-3 & \\
04 & 03:52:40.31 & --00:05:23.3 & 95-015 & & \\
05 & 03:52:40.62 & --00:05:06.1 & 95-016 & & \\
06 & 03:52:41.17 & ~~00:31:20.7 & 95-301 & & SA95-301 \\
07 & 03:52:42.17 & ~~00:31:17.1 & 95-302 & & \\
08 & 03:52:44.82 & --00:03:34.1 & & 3-4 & \\
09 & 03:52:53.10 & --00:03:52.4 & & 3-5 & \\
10 & 03:52:54.18 & ~~00:00:18.4 & 95-096 & & SA95-096 \\
11 & 03:52:54.87 & --00:03:16.1 & & 3-6 & \\
12 & 03:52:57.50 & --00:00:19.8 & 95-097 & 3-7 & \\
13 & 03:53:00.23 & ~~00:02:48.5 & 95-098 & & \\
14 & 03:53:00.78 & ~~00:00:15.2 & 95-100 & & \\
15 & 03:53:04.14 & ~~00:02:49.5 & 95-101 & & \\
16 & 03:53:07.58 & ~~00:01:10.3 & 95-102 & & \\
17 & 03:53:10.62 & ~~00:27:22.3 & 95-252 & & \\
18 & 03:53:13.24 & ~~00:16:22.2 & 95-190 & & \\
19 & 03:53:20.59 & ~~00:16:34.3 & 95-193 & & \\
20 & 03:53:21.32 & --00:00:19.0 & 95-105 & & \\
21 & 03:53:25.18 & ~~00:01:22.2 & 95-106 & & \\
22 & 03:53:25.65 & ~~00:02:19.8 & 95-107 & & \\
23 & 03:53:40.10 & --00:01:11.9 & 95-112 & & \\
24 & 03:53:41.19 & --00:02:32.8 & 95-041 & & \\
25 & 03:53:43.66 & --00:04:33.9 & 95-042 & & \\
26 & 03:53:44.17 & ~~00:29:49.5 & 95-317 & & \\
27 & 03:53:46.98 & ~~00:26:40.4 & 95-263 & & \\
28 & 03:53:47.80 & --00:00:47.8 & 95-115 & & \\
29 & 03:53:48.59 & --00:03:02.0 & 95-043 & & \\
30 & 03:54:16.28 & ~~00:18:51.8 & 95-271 & & \\
31 & 03:54:19.45 & ~~00:36:31.3 & 95-328 & & \\
32 & 03:54:23.89 & ~~00:37:06.7 & 95-329 & & \\
33 & 03:54:30.73 & ~~00:29:03.5 & 95-330 & & \\
34 & 03:54:44.23 & ~~00:27:19.2 & 95-275 & & \\
35 & 03:54:45.86 & ~~00:25:53.3 & 95-276 & & \\
            \hline
         \end{tabular}
   \end{table}
\setcounter{table}{3}
   \begin{table}
   \caption{Continued}
         \begin{tabular}{cccccc}\hline
           \hline
 Number & $\alpha$(2000.0) & $\delta$(2000.0) & L-1992 & G-2000  & L-1983 \\
  (1) & (2)  & (3) & (4) & (5) & (6) \\
           \hline
36 & 03:54:49.52 & --00:07:04.4 & 95-060 & & \\
37 & 03:54:49.93 & ~~00:10:08.0 & 95-218 & & \\
38 & 03:54:51.66 & ~~00:05:21.1 & 95-132 & & SA95-132 \\
39 & 03:55:00.39 & --00:02:54.5 & 95- 62 & & \\
40 & 03:55:03.72 & ~~00:03:26.4 & 95-137 & & \\
41 & 03:55:04.65 & ~~00:03:07.6 & 95-139 & & \\
42 & 03:55:08.73 & ~~00:14:34.0 & 95-227 & & \\
43 & 03:55:09.37 & ~~00:01:20.2 & 95-142 & & \\
44 & 03:55:31.14 & --00:09:13.8 & 95-074 & & SA95-074 \\
45 & 03:55:38.81 & ~~00:10:43.0 & 95-231 & & \\
46 & 03:55:41.54 & ~~00:26:36.9 & 95-284 & & \\
47 & 03:55:44.10 & ~~00:25:09.6 & 95-285 & & \\
48 & 03:55:44.42 & ~~00:07:02.3 & 95-149 & & \\
            \hline
         \end{tabular}
   \end{table}
{\tiny
   \begin{table}
   \caption{The SED of the Landolt standard stars. 
            The value of 00.00 means that there is no measurement 
            because the star is saturated in this 
            band. The magnitude error of such star is 
            estimated to be 0.03.}
         \begin{tabular}{cccccccccccccccc}\hline
           \hline
N. &  $a$   &  $b$  &  $c$  &  $d$  &  $e$  &  $f$  &  $g$  &  $h$  & 
  $i$  &  $j$  &  $k$  &  $m$  &  $n$  &  $o$  &  $p$  \\
 (1) & (2)  & (3) & (4) & (5) & (6) & (7) & (8) & (9) & (10) & (11) & (12) & (13) & (14) & (15) & (16) \\
           \hline
01&17.39&16.54&16.24&15.86&15.66&15.52&15.29&15.23&15.12&15.06&15.04&14.95&14.95&14.89&14.88\\
02&17.45&16.66&16.07&15.50&15.23&15.05&14.74&14.69&14.55&14.49&14.45&14.33&14.31&14.23&14.24\\
03&15.85&15.08&14.77&14.44&14.21&14.12&13.91&13.88&13.79&13.74&13.73&13.61&13.62&13.56&13.58\\
04&13.30&12.57&12.13&11.71&00.00&00.00&00.00&00.00&00.00&00.00&00.00&00.00&00.00&10.81&10.83\\
05&17.92&16.66&16.03&15.05&14.72&14.56&13.97&13.82&13.58&13.48&13.34&13.16&13.13&13.00&12.96\\
06&15.23&13.85&13.06&12.15&00.00&11.46&11.09&00.00&00.00&00.00&00.00&00.00&00.00&00.00&10.00\\
07&14.20&13.33&12.83&12.21&00.00&00.00&00.00&00.00&00.00&00.00&00.00&00.00&00.00&11.08&11.08\\
08&18.42&17.61&16.99&16.41&16.07&15.96&15.65&15.57&15.41&15.35&15.29&15.18&15.17&15.07&15.05\\
09&16.61&15.72&15.43&15.17&14.94&14.85&14.64&14.60&14.50&14.42&14.41&14.32&14.33&14.26&14.26\\
10&11.49&10.22&10.25&10.36&00.00&00.00&00.00&00.00&00.00&00.00&00.00&00.00&00.00&10.17&10.21\\
11&16.15&15.35&14.89&14.53&14.22&14.13&13.88&13.83&13.70&13.62&13.58&13.48&13.49&13.41&13.41\\
12&17.25&16.31&15.86&15.40&15.10&14.96&14.65&14.58&14.42&14.35&14.29&14.15&14.14&14.01&14.02\\
13&17.68&16.93&16.03&15.17&14.84&14.72&14.18&14.05&13.85&13.75&13.63&13.49&13.48&13.27&13.34\\
14&17.66&16.90&16.49&16.10&15.79&15.70&15.44&15.37&15.23&15.18&15.11&14.99&14.98&14.86&14.91\\
15&14.79&14.09&13.57&13.14&12.89&12.77&12.52&00.00&00.00&00.00&00.00&12.17&12.18&12.03&12.12\\
16&17.87&17.18&16.68&16.23&15.93&15.78&15.48&15.40&15.24&15.18&15.12&14.98&14.95&14.80&14.85\\
17&19.59&18.17&17.38&16.44&15.91&15.67&15.13&14.94&14.67&14.52&14.32&14.20&14.11&13.93&13.86\\
18&14.46&13.02&12.91&12.74&00.00&00.00&00.00&00.00&00.00&00.00&00.00&12.57&12.60&12.51&12.47\\
19&17.83&16.91&16.05&15.09&14.71&14.61&14.04&13.89&00.00&13.58&13.43&13.33&13.28&13.19&13.12\\
20&16.31&15.56&14.82&14.18&13.85&13.74&13.38&13.29&00.00&13.05&12.97&12.88&12.86&12.71&12.75\\
21&18.06&17.13&16.49&15.90&15.56&15.35&15.00&14.88&14.68&14.61&14.53&14.39&14.33&14.15&14.21\\
22&19.84&19.02&18.29&17.28&16.91&16.65&16.02&15.88&15.50&15.29&14.94&14.85&14.75&14.43&14.45\\
23&17.82&17.01&16.61&16.12&15.77&15.60&15.29&15.21&15.00&14.91&14.81&14.70&14.65&14.51&14.54\\
24&16.50&15.66&15.16&14.70&14.35&14.18&13.86&13.76&13.56&13.46&13.36&13.25&13.19&13.02&13.07\\
25&15.09&15.14&15.29&15.35&15.39&15.51&15.70&15.79&15.98&16.02&16.16&16.26&16.39&16.32&16.68\\
26&17.23&16.03&15.22&14.41&14.02&13.69&13.19&13.01&00.00&12.66&00.00&12.34&12.25&12.16&12.05\\
27&17.18&15.73&14.72&13.75&13.19&12.95&12.39&12.20&11.93&00.00&00.00&00.00&11.43&11.29&11.18\\
28&17.18&16.37&15.85&15.36&15.01&14.82&14.50&14.42&14.21&14.13&14.02&13.91&13.84&13.75&13.74\\
29&12.37&11.61&11.33&11.25&00.00&00.00&00.00&00.00&00.00&00.00&00.00&00.00&00.00&10.49&10.57\\
30&17.07&16.02&15.24&14.59&14.14&13.87&13.44&13.30&00.00&00.00&12.77&12.59&12.50&12.37&12.27\\
31&17.70&16.39&15.53&14.67&14.14&13.78&13.22&13.02&12.66&00.00&00.00&12.08&11.98&11.78&11.65\\
32&17.98&17.17&16.28&15.37&15.06&14.89&14.32&14.15&13.91&13.83&13.65&13.52&13.49&13.36&13.29\\
33&17.62&16.28&14.94&13.72&00.00&12.56&00.00&00.00&00.00&00.00&00.00&00.00&00.00&00.00&09.66\\
34&18.33&16.88&15.78&14.78&14.18&13.79&13.10&00.00&00.00&00.00&00.00&00.00&11.72&11.52&11.36\\
35&17.69&16.73&15.79&14.90&14.54&14.40&13.84&13.68&00.00&13.37&13.19&13.06&13.02&12.92&12.83\\
            \hline
         \end{tabular}
   \end{table}
\setcounter{table}{4}
   \begin{table}
   \caption{Continued}
         \begin{tabular}{cccccccccccccccc}\hline
           \hline
N. &  $a$   &  $b$  &  $c$  &  $d$  &  $e$  &  $f$  &  $g$  &  $h$  & 
  $i$  &  $j$  &  $k$  &  $m$  &  $n$  &  $o$  &  $p$  \\
  (1) & (2)  & (3) & (4) & (5) & (6) & (7) & (8) & (9) & (10) & (11) & (12) & (13) & (14) & (15) & (16) \\
           \hline
36&15.48&14.83&14.38&13.91&13.55&13.52&13.26&13.20&00.00&13.00&12.92&12.87&12.83&12.76&12.83\\
37&14.11&13.35&12.88&12.51&00.00&00.00&00.00&00.00&00.00&00.00&00.00&00.00&11.68&11.65&11.63\\
38&14.09&12.79&12.51&12.29&00.00&00.00&00.00&00.00&00.00&00.00&00.00&00.00&11.90&11.82&11.82\\
39&17.37&16.26&15.41&14.50&14.00&13.76&13.27&13.15&12.85&00.00&12.61&12.49&12.40&12.23&12.22\\
40&18.17&17.20&16.50&15.45&15.06&14.82&14.17&13.99&13.69&13.53&13.25&13.11&13.01&12.83&12.83\\
41&14.88&14.17&13.47&12.75&00.00&00.00&00.00&00.00&00.00&00.00&00.00&00.00&11.52&11.41&11.45\\
42&17.65&16.86&16.63&16.32&16.05&15.88&15.63&15.53&15.36&15.28&15.19&15.07&15.01&14.89&14.89\\
43&14.75&13.85&13.63&13.29&13.06&13.00&12.81&00.00&00.00&00.00&12.68&12.57&12.54&12.47&12.51\\
44&14.51&13.61&13.10&12.32&00.00&00.00&00.00&00.00&00.00&00.00&00.00&00.00&00.00&10.59&10.56\\
45&16.18&14.96&14.75&14.46&14.29&14.25&14.14&14.11&14.10&14.04&14.03&14.00&13.99&13.87&13.94\\
46&17.45&16.35&15.53&14.69&14.19&13.92&13.38&13.22&00.00&00.00&00.00&12.41&12.35&12.10&12.07\\
47&17.96&16.95&16.65&16.22&15.86&15.68&15.35&15.24&15.05&14.98&14.81&14.68&14.66&14.51&14.46\\
48&15.44&14.09&13.13&12.12&00.00&00.00&00.00&00.00&00.00&00.00&00.00&00.00&00.00&00.00&00.00\\
            \hline
         \end{tabular}
   \end{table}
\end{document}